\documentclass[12pt]{article}

\usepackage{graphicx}

\begin{document}

\title{A Quantum Many-Body Instability in the Thermodynamic Limit}

\author{Marco Frasca \\
        Via Erasmo Gattamelata, 3 \\
        00176 Roma (Italy)}

\date{\today}

\maketitle

\abstract{
Intrinsic decoherence in the thermodynamic limit is shown for a large class of many-body
quantum systems in the unitary evolution in NMR and cavity QED. The effect largely depends 
on the inability of the system to recover the phases. Gaussian decaying in time 
of the fidelity is proved for spin systems and radiation-matter interaction.
}


\newpage

Recent experimental findings \cite{pasta0,pasta1,pasta2,pasta3,pasta4} in NMR with
organic molecular crystals have shown that an intrinsic decoherence effect seems present
in a many-body system in the thermodynamic limit appearing in the form of a Gaussian decay
in time depending on the coupling between spins and their number. 
These experiments, that are known just
in the NMR community, give a strong support to an understanding of decoherence as an intrinsic effect
arising from the thermodynamic limit applied to unitary evolution \cite{fra3,fra4,fra5,fra6,fra7,fra8,fra9}.
These latter theoretical results are based on a concept of singular limit in time. Indeed, 
the existence of such an effect would
give a relevant answer both to the measurement problem and the question of the irreversibility:
both are essential to the understanding of a classical world emerging from quantum mechanics.

Suter and Krojanski showed again, in
a recent NMR experiment \cite{suter}, decoherence with a Gaussian decay in time but was
definitely proved that the time-scale of the decay depends on the square root of the number of spins,
confirming and extending previous experiments. These results seem to defeat ordinary understanding
of environmental decoherence.

Besides, recent experiments with cavities realized by Haroche's group \cite{har1,har2} have
produced asymptotic states with a large number of photons as foreseen by Gea-Banacloche \cite{gb1,gb2}
and further analyzed by Knight and Phoenix \cite{kn1,kn2}. As firstly pointed out by Gea-Banacloche,
these states support a view of quantum measurement described by decoherence in the thermodynamic
limit, in agreement with the view above. In this case one has that the thermodynamic limit makes the
revival time very large and one is left with an apparently collapsed wave function.

A singular limit in time appears when an oscillating function has the frequency going to
infinity. In this case, sampling the function to recover it becomes increasingly difficult.
In the thermodynamic limit, with the frequency directly proportional to the number of particles,
it becomes very easy to reach oscillations with a period of the order of Planck time where
physics is expected to change \cite{rov} and there is no possibility at all to sampling
a periodic function. So, as expected for too rapidly oscillating functions, the physical
result one gets is an average. This gives a very simple method to generate random numbers
with a sin function (see fig.1).

We will give a rather general result on the existence of an intrinsic decoherence effect
in an interacting many-body system. The result will rely both
on a theorem recently proved \cite{hmh}, showing
Gaussian decay in the thermodynamic limit for many-body systems in the fidelity, that we will
extend to the case of radiation-matter interaction.

\begin{figure}[t,b,p]
\begin{center}
\includegraphics[height=1\textwidth,width=1\textwidth]{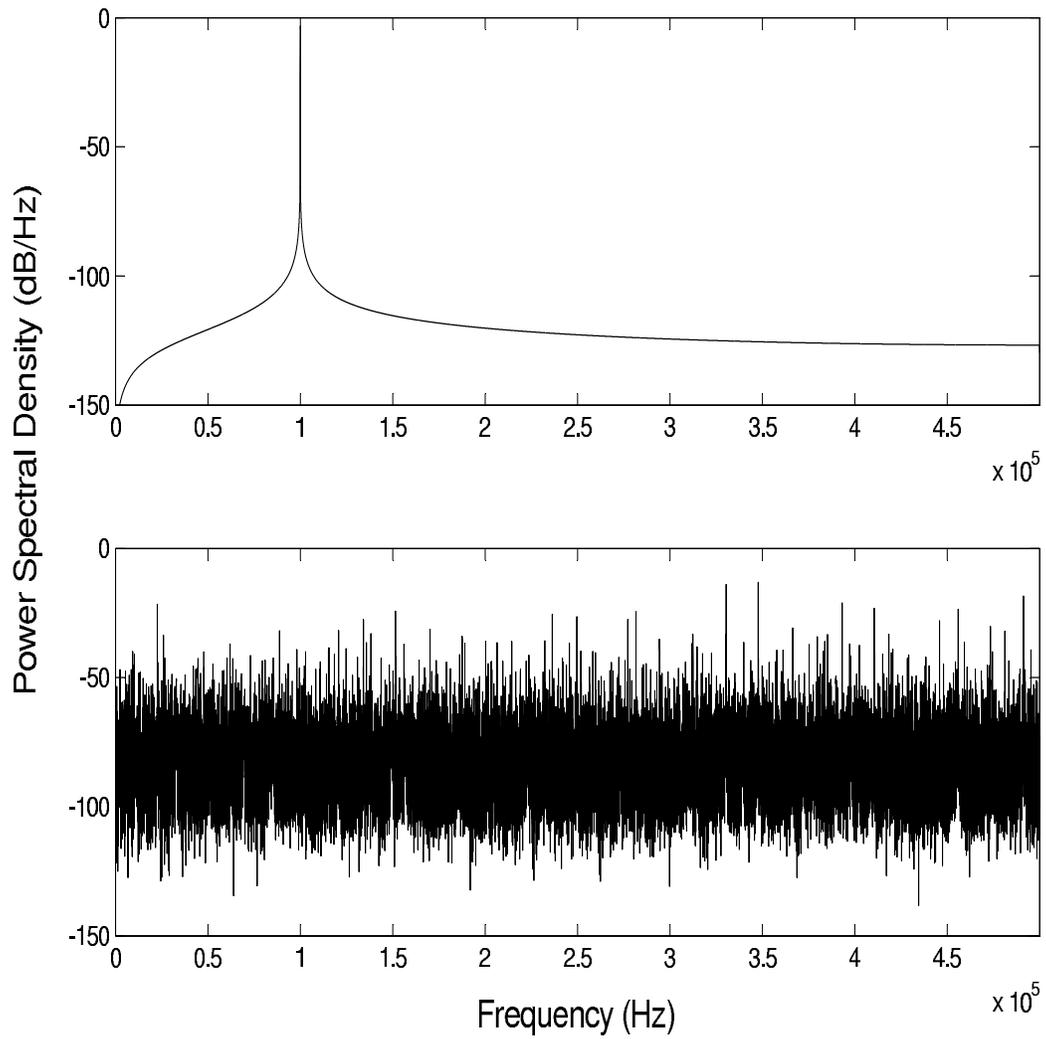}
\caption{\label{fig:fig1} Periodograms of the function $\sin(2\pi N t)$ with $N=10^5$ Hz (upper) 
and $N=10^{43}$ Hz (lower) both sampled with a frequency of 1 MHz.}
\end{center}
\end{figure}

A non-interacting many-body system can already display, for a large class of states, a set of observables
that behave classically. This result was proven in Ref.\cite{fra5}. We give here an overview
of this result. One consider a Hamiltonian $H = \sum_{i=1}^N H_i$ and a set of 
single particle states $|\psi_i\rangle$ that are not eigenstates of $H_i$. We can form
an initial product state $|\psi(0)\rangle = \prod_{i=1}^N|\psi_i\rangle$ that is not an eigenstate
of $H$. It is easy to show that the Hamiltonian behaves as a classical observable with respect
to the state $|\psi(0)\rangle$ as the mean value $\langle H\rangle\propto N$ and the squared
variance $\langle H^2\rangle-\langle H\rangle^2\propto N$, so, in the thermodynamic limit
we have proved that an observable behaving classically does exist being the quantum fluctuation
negligible small with respect to the mean value in the large $N$ limit. The same argument applies
for any other observable commuting with $H$ that, in this way, does not evolve in time. This
result can be extended to a non-commuting observable $A$ defined like the Hamiltonian $A=\sum_{i=1}^N A_i$ 
by evolving it in time with the Heisenberg equations of motion \cite{fra5}. Again one has
$\langle A\rangle\propto N$ and $(\Delta A)^2=\langle A^2\rangle-\langle A\rangle^2\propto N$
and quantum fluctuations are negligible small with respect to the mean value. We note
that we have just obtained our result for extensive observables, that is, we consider operators
that increase like the number of the systems. This point is crucial but does not limit at
all the generality of our results.

Already at this stage we can see a deep conceptual similarity between statistical mechanics
and quantum mechanics for non-interacting systems. Indeed, our aim is to show how the
thermodynamic limit too has a central role in quantum mechanics beyond the formal
Wick rotation on the evolution operator. One could
think to extend the above result to interacting systems by a cluster expansion\cite{kada} but we
do not pursue this matter here. We would like to see how extensive observables are able
to produce decoherence intrinsically on a quantum system. This is enough to prove the
very existence of such an effect.

Indeed, one can prove that the unitary evolution in the thermodynamic limit, for
spin systems and radiation-matter interaction, when ruled by the Dicke model \cite{dic}, is
Gaussian and the decay time-scale is given by the inverse of the variance 
of the Hamiltonian on the initial state.
We note here that this behavior is common at small times for any quantum system 
as proved by Misra and Sudarshan \cite{ms}. Indeed, this is due to the fact that the
evolution, in the thermodynamic limit, gets significant contributions only at very
small times and goes rapidly to zero otherwise. Rather, recurrence can happen and
we will see how this effect is not really observable in our case being the
variance very large, increasing with the number of systems, and making in
this way the duration of a recurrence too short.

In order to prove that a spin system undergoes Gaussian decay, we need the following
theorem due to Hartmann, Mahler and Hess (HMH)\cite{hmh}. This theorem can be stated in the 
following way:

{\bf Theorem (HMH)}
{\sl
If the many-body Hamiltonian $H$ and the product state $|\phi\rangle$ satisfy
\begin{equation}
\label{eq:cond}
     \sigma_\phi^2\ge NC
\end{equation}
for all $N$ and a constant $C$ and if each $H_i$ is bounded as $\langle\chi|H_i|\chi\rangle\le C'$
for all normalized states $|\chi\rangle$ and some constant $C'$, then, for fidelity, the following holds
\begin{equation}
    \lim_{N\rightarrow\infty}|\langle\phi|e^{-iHt}|\phi\rangle|^2=e^{-\sigma_\phi^2 t^2}.
\end{equation}
}

We have put $\sigma_\phi^2=\langle H^2\rangle - \langle H\rangle^2$ 
and the hypothesis of nearest neighbor interactions applies. 
The decay depends crucially on the properties of the Hamiltonian that rules
quantum dynamics. Particularly we have a dependence on the number of systems $N$  and on the
coupling between systems. For spin systems with nearest neighbor interaction  
one has $\sigma_\phi\propto\sqrt{N}$ for any state and
a Gaussian decay is indeed observed differently from the customary exponential decay observed
for environmental decoherence. We will get an analogous result for the Dicke model. It is
interesting to note that the HMH theorem, in the light of the theorem of Misra and
Sudarshan on quantum evolution is just saying that, for a spin system in the thermodynamic limit,
what is really relevant is the evolution at small times going otherwise rapidly to zero.
About the recurrence, it should be said that the duration of this effect is really
too short to be observed. We will make this matter more evident in the case of the
Dicke model.

The HMH theorem gives an explanation of the results obtained
in NMR experiments so far\cite{pasta0,pasta1,pasta2,pasta3,pasta4,suter}. 
Decoherence can be seen as an
intrinsic effect of quantum dynamics in the thermodynamic limit 
due to the inability of a many-body system to maintain
its coherence during time evolution. The information on the phases is unavoidably lost. As a
by-product we have shown that irreversibility in macroscopic dynamics, imposed by Boltzmann
through the hypothesis of molecular chaos, finds its natural explanation in quantum dynamics.
The Hamiltonian applied to these experiments was a nearest neighbor spin-spin one
as those considered in the HMH theorem. So, as shown
we should expect a Gaussian decay with $\sigma_\phi\propto\sqrt{N}d$ being $d$
a function of the couplings between spins in agreement with experimental results 
as also beautifully confirmed by Krojanski and Suter\cite{suter}.

In radiation-matter interaction at low energies several approximations are generally made
that proved to hold as the field of quantum optics is there to testify. Indeed, one
can assume that atoms have just two levels, the dipole approximations and that
only the linear term in the radiation matter interaction (non-relativistic approximation)
should be kept.
The most general Hamiltonian that holds in this case is given by the Dicke model. When
also the rotating wave approximation is applied the rather ubiquitous Jaynes-Cummings model is
recovered. The rather wide applicability of this model can be seen also for
arrays of Josephson junctions in a cavity \cite{frak}, 
that could be used in experiments to observe this kind
of intrinsic decoherence. 

The Dicke model have had recently several analysis displaying a rich dynamics
in the thermodynamic limit \cite{fra6,fra7,fra9,eb1,eb2}. Here we would like to see how
coherence decays while time evolves. The Hamiltonian of the Dicke model can be written as
\begin{equation}
    H = \frac{\Delta}{2}\sum_{n=1}^N\sigma_{zi}+\omega a^\dagger a +
	g\sum_{n=1}^N\sigma_{xi}(a^\dagger + a)
\end{equation}
being $\Delta$ the separation between the levels of the two-level atoms,
$g$ the coupling constant, $N$ the number of two-level atoms, $a^\dagger$, $a$
the creation and annihilation operators for the radiation mode, $\sigma_{xi}$,
$\sigma_{zi}$ Pauli spin matrices for the i-th atom. As proved in Ref.\cite{fra9}, when
the coupling constant is kept fixed and the number of particles goes to infinity the model
becomes integrable and described by the Hamiltonian
\begin{equation}
    H_F = \omega a^\dagger a + g\sum_{n=1}^N\sigma_{xi}(a^\dagger + a)
\end{equation}
with $x$ the proper quantization axis. Then, the time evolution operator can be written as 
\begin{equation}
    U_F(t)=e^{-iH_Ft}=e^{i\hat\xi(t)}e^{-i\omega a^\dagger at}\exp[\hat\alpha(t)a^\dagger-\hat\alpha^*(t)a]
\end{equation}
being
\begin{equation}
    \hat\xi(t)=\frac{\left(\sum_{i=1}^N\sigma_{xi}\right)^2g^2}{\omega^2}(\omega t-\sin(\omega t))
\end{equation}
and
\begin{equation}
    \hat\alpha(t)=\frac{\left(\sum_{i=1}^N\sigma_{xi}\right)g}{\omega}(1-e^{i\omega t}).
\end{equation}
As for decoherence in the thermodynamic limit the preparation of the system is critical, and
this is due to the general property of coherence in quantum systems \cite{fracbm}, we assume
the two-level systems initially in a state having a large part of atoms all in the same
state. To make computation simpler we can directly assume the ground state where all the atoms
are in their ground state as this cannot change our conclusions. In this way we have
just to consider unitary evolution for a generic radiation state $|\chi\rangle$ given by 
\begin{equation}
    U_R(t)=e^{i\xi(t)}e^{-i\omega a^\dagger at}\exp[\alpha(t)a^\dagger-\alpha^*(t)a]
\end{equation}
and now we have
\begin{equation}
    \xi(t)=\frac{N^2g^2}{\omega^2}(\omega t-\sin(\omega t))
\end{equation}
and
\begin{equation}
    \alpha(t)=\frac{Ng}{\omega}(1-e^{i\omega t}).
\end{equation}
The behavior of this evolution for $N\rightarrow\infty$ for the state $|\chi\rangle$ is
what interests us. So we do the following expansion
\begin{equation}
     |\chi\rangle = \sum_{n}c_n|n\rangle
\end{equation}
being $|n\rangle$ radiation number states. It is straightforward to obtain the
following expansion for the decaying amplitude
\begin{eqnarray}
\label{eq:sum}
    \langle\chi|U_R(t)|\chi\rangle &=& e^{i\xi(t)}
	e^{-\frac{N^2g^2}{\omega^2}(1-\cos(\omega t))}
	\sum_{m,n}c_m^*c_ne^{-im\omega t}\frac{n!}{m!}\times \\ \nonumber
	& &\left[\frac{Ng}{\omega}(1-e^{i\omega t})\right]^{m-n}
	L^{m-n}_n\left[\frac{2N^2g^2}{\omega^2}(1-\cos(\omega t))\right]
\end{eqnarray}
being $L^{m-n}_n(x)$ the associated Laguerre polynomials and
use has been made of the relation \cite{kn3}
\begin{equation}
   \langle m|\exp[\alpha a^\dagger-\alpha^* a]|n\rangle=
   e^{-\frac{1}{2}|\alpha|^2}\frac{n!}{m!}\alpha^{m-n}
   L^{m-n}_n(|\alpha|^2).
\end{equation}
From a strictly numerical standpoint we note that the exponential
$e^{-\frac{N^2g^2}{\omega^2}(1-\cos(\omega t))}$, a periodic
function, has values sensibly different from zero only for small times
in the limit $N\rightarrow\infty$, otherwise it goes rapidly to zero and does
this through a Gaussian. So, we can represent this function as a periodical
reappearing of a Gaussian decaying function,
$e^{-\frac{N^2g^2t^2}{2}}$, 
with the period being given by $2\pi/\omega$.
This fact drives all the unitary evolution as the sum in eq.(\ref{eq:sum}) is
different from zero only when the decaying function is significantly different from
zero, that is, at small times. This in turns means that, taking the thermodynamic limit
on the Dicke model implies that the time evolution is meaningful just for small times.
We can apply straightforwardly the theorem by Misra and Sudarshan and we can conclude that
the square of the decaying amplitude (\ref{eq:sum}) is given by a Gaussian, that is,
\begin{equation}
     \lim_{N\rightarrow\infty}|\langle\chi|U_R|\chi\rangle|^2=e^{-\sigma_H^2 t^2}
\end{equation}
being $\sigma_H^2=\langle H^2\rangle - \langle H\rangle^2$ with $H$ the Dicke Hamiltonian.
This extends the HMH theorem to the radiation-matter interaction for the case of the
Dicke Hamiltonian. 

In order to see the above result at work, we apply it to some known radiation states.
For a Fock state $|n\rangle$ it is easy to verify that $\sigma_\chi=\sqrt{2n+1}Ng$. Similarly,
one could get $\sigma_\chi=\sqrt{nN^2g^2+\omega^2/4}$ for a superposition of Fock states as 
$|\chi\rangle=[|n\rangle+|n-1\rangle]/\sqrt{2}$. 
One can see this same results obtained directly from the unitary evolution 
that for the ground state gives
\begin{equation}
    F_0=e^{-2\frac{N^2g^2}{\omega^2}[1-\cos(\omega t)]}.
\end{equation}
that as said above reduces to 
\begin{equation}
   F_0\approx e^{-N^2g^2t^2}.
\end{equation}
On the same ground we can see that, for a different Fock state one has, at very small times,
\begin{equation}
    F_n\approx e^{-N^2g^2t^2} L_n^2(N^2g^2t^2)
\end{equation}
being $L_n$ the Laguerre polynomial of n-th order. This gives a rather good agreement with
our result pointed out above $F_n=e^{-(2n+1)N^2g^2t^2}$
as, at very small times, $L_n^2(N^2g^2t^2)\approx e^{-2nN^2g^2t^2}$. The
same procedure yields our result above for a superposition of Fock states. 

So, a Gaussian decay should be observed in the thermodynamic limit for the Dicke model on superposition
states of the radiation field. This result is very important in view of the fact that
measurement in quantum mechanics is realized through radiation-matter interaction.
Finally, this result is in agreement both with
the numerical results presented in Ref.\cite{eb1,eb2} and theoretical ones given in
Ref.\cite{fra6,fra7,fra9}. 

This Gaussian decay is
recurrent with a period $2\pi/\omega$ but, in the thermodynamic limit,
the possibility to observe such a recurrence becomes increasingly difficult due to the
increasingly smallness of the time scale of the Gaussian decay, that goes like $1/N$,
granting an even smaller width of the Gaussian, that is the duration of the recurrence.
If $N$ is large enough the duration of the recurrence assumes unphysical values.

It is rather interesting to see how much should be the infinity in the thermodynamic limit
to get physically sensible results. Indeed, we learned from phase transitions, through the
Onsager solution and the theorems of Yang and Lee \cite{ons,yl} that require the thermodynamic
limit as the partition function has no zeros on the real axis, that Avogadro number is
indeed enough for a physically sensible result and this is our everyday experience. The same
can be said for our case where very small times are involved increasing the number of particles.

In conclusion, we have given here a significant result for intrinsic decoherence in the thermodynamic
limit for the unitary evolution in many-body quantum mechanics
amenable to experimental tests. On this basis, we have seen that both spin systems and
the Dicke model that describes radiation-matter interaction for an ensemble of two-level atoms 
is not able to keep coherence in the thermodynamic limit.

\end{document}